RESEARCH

# New Market Creation via Innovation: A Study on Tata Nano

Dr. Swati Singh & Dr. Manoj Joshi*

**ABSTRACT**

***Objective of the paper-*** *This research paper focuses on how innovations support new market creation emerging from latent opportunities for low-income group. It also emphasizes on novel strategies that can be implemented for sustaining. The paper concludes with a discussion on the implications of the study and directions to stimulate future research on the subject.*

***Design/methodology/approach-*** *A hybrid methodology comprising of a case study on the 2200 USD (equivalent to INR 1,00,000) car Tata Nano and supported with a quantitative survey on a random sample of prospective Nano customers has been deployed. The qualitative case study helps to understand the macro picture on how innovation helped create new market for the masses. The quantitative survey measures the impact of various attributes of Nano on customer's intent to purchase. These attributes have been identified through interviews of prospective Nano customers. Factor analysis has been deployed to establish corporate brand equity, aesthetics, and value for money & reliability as main attributes of Nano. Regression analysis is used to find out the influence of these attributes on intent to purchase Nano.*

***Findings*** *The study highlights the power of brand equity in new market creation and n innovation sustainability. The survey indicates the perception of prospective customers about Nano and how it impacts their buying decisions. The case reflects on how companies can successfully create products and at the same time positions it for its market adoption. The paper concludes with the development of new strategies to sustain this innovation.*

***Originality/value*** *The study's originality springs from the diffusion of innovation theories and emphasising the need for organizations targeting the bottom of the pyramid. The case study reflects how a dream became a reality to own an affordable four-wheeler by masses.*

***Implications*** *A key implication of this paper is that the organizations must adopt 'blue ocean strategy' (Kim & Mauborgne, 2005) whereby organizations can generate high growths and profits by creating new demand in an uncontested market space. It also provides insights into understanding market behaviour towards innovation and adoption of low cost products.*

**Dr. Swati Singh** Assistant Professor, Marketing, Amity Business School, Amity University, Uttar Pradesh, Lucknow.
**Email:** swatirastogisingh@gmail.com
**Dr. Manoj Joshi,** FIE (Mech.Engg.), Ph.D (Strategy) Professor of Strategy, Entrepreneurship and Innovation, Amity Business School, Amity University, Lucknow.
**Email:** manoj.m.joshi@gmail.com

***Directions for future Research-*** *Similar studies can be carried out that can help in comparative analysis of different products targeted at the bottom of the pyramid. There is a possibility that selling to the bottom of the pyramid could be attractive in some industries or firms but not others. It would provide better understanding of what features help in sustainability of such products in the long run. It would also be interesting to find out how such products would be helpful in reverse innovation.*

***Research limitations-*** *This case focuses on an automobile, which is a high involvement product. The implications and strategies could differ for products in other categories. The sample size for the study is not very large so it only gives an indication and not generalization of the findings. The case study on Nano focuses on the marketing and strategy perspective rather than production and technology.*

**KEY WORDS:** Innovation, New Market Creation, Strategy, Nano, India
**Type of Research Paper** Mixed Method Research comprising of case study and empirical survey

## INTRODUCTION

Multinational organizations are facing tough competition within global markets. Developed western markets are stagnant and competitive with less scope for creation and adoption of innovations. This has forced them for geographical extension and to seek opportunities in emerging economies and markets like India, China, Indonesia, Brazil etc. Both multinationals and home-grown organizations are attracted towards rapidly growing markets to look for better opportunities in untapped emerging economies (Munshi, 2011). This results in a 'red ocean' (Kim & Mauborgne, 2005) and limits the scope for geographical extension in these economies (Roy, 2010).

In India, the offerings from multinational organizations majorly focus on the upper class or upper middle class segments who easily adapt these products, originally meant for developed countries. The local organizations imitate these offerings without investing their resources in recognizing the differences in these markets. The lower segments finally get the trimmed down, no frills versions of the same product without the organizations understanding the difference in the needs of these segments. These products rarely satisfy the customers as they were originally created for a different set of consumer base. Hence, the middle class and lower middle class markets crave for their own niche offerings, designed specifically for them.

India needs innovations at the grass root level. This strata comprising of bottom of the pyramid (Prahalad, 2004) has a tremendous potential for new market creation as there is untapped latent potential of its consumers. This bottom of the pyramid is very lucrative as it comprises of huge chunk of its total population (Hammond et al., 2008), increasing consumer awareness, new buying patterns, growth in income as well as literacy and low competition (Van den Waeyenberg and Luc Hens, 2008). There is a tussle between organizations to satisfy this huge segment. This has led to creation of 'value for money' products by these companies (Bellman, 2009). Some prime examples of products designed for Indian market are, Nokia's sturdy

'Made in India phone' with longer battery life which is suitable for Indian conditions like frequent power cuts, General Electric's has designed cheaper, stripped-down electrocardiogram (ECG) version at 1,000 USD and 'The Chottu Cool' refrigerator by Voltas for the rural market. The biggest landmark of such an innovation is Tata's Nano car priced at 2,200 USD.

## LITERATURE REVIEW AND RESEARCH HYPOTHESIS

Innovation represents resultant change to overcome saturation of the market and to gain and retain competitiveness in today's cut-throat competition. Tushman and Nadler (1986), defined innovation as process that generate creative products services or production processes for a particular business. Innovations can spring in any form, be it new form customers, markets, delivery or and products (Joshi,2007; 2010). In order to succeed with innovative products organizations need to have customer trust in their brand. This creates strong bonds and positive attitude towards products. Organizations expertise, reliability and intentions help in winning customer trust. This brings customer confidence in understanding of the product offering. There is direct relationship between trust and brand equity (Morgan & Hunt, 1994; Garbarino and Johnson, 1999). It brings forth commitment of customers in the organization. This creates the strong bond between customer and organization leading to easy acceptability of the new offerings by the organization.

Identification of these needs leads to new market creation with offerings designed especially for them, subjected to availability of local resources. Organizations are now focusing on this unexplored market segment, specifically called the bottom of the pyramid. These innovative products are tailor-made with an affordable price in mind (Vijayraghavan, 2010). These products stand apart not just for their functionality but also in their ingenious use of local commodities and ability to compete with more technologically advanced products of the west. These offerings eventually find market in the middle class also (Parthasarathy et al., 2010).

It is imperative for the organizations to adapt the marketing approach to meet the characteristics of these consumers (Pitta et al. 2008). Key to success for organisations is to have an opportunity seeking behaviour in order to identify and understand these markets. These novel products require organizations to reinvent their business models with customized marketing and positioning strategies to woo the customers. To survive in this new market segment they should signify the relevance and utility of these products to the target customers.

Major strategy while initiating innovation for bottom of pyramid is to understand cost parameters. In order to achieve affordability, they must reduce the costs of production and simplify the products (Ramaswamy and Schiphorst, 2000). It involves step by step planning from need identification to manufacturing of the end product. The innovation arises out of best utilization of available resources with price-sensitivity as its major driver. The organisations looking for latent opportunities have to devise strategies to not only excel but also to gain knowledge, insights and understanding of the new market segments, that greatly enhance their abilities to compete (Wood et al., 2008). They should expand the reach of stakeholders, choose non-traditional partners, and adopt a native capability.

The objective of the study is to explore how innovation supports new market creation emerging from identification of latent opportunities for low-income group and what novel strategies can be implemented to sustain it. It also attempts to find out which attributes of Nano acts as deciding forces for customers to purchase Nano.

## RESEARCH DESIGN

This research entails a mixed methodology comprising of a case study on the 2,200 USD (equivalent to INR 1,00,000) car Tata Nano supported with a quantitative survey on a random sample of prospective Nano customers. The qualitative case study understand the macro picture of how innovation led to new market creation by identifying the latent opportunities in the low income group of India (Singh & Srivastava, 2012). This is coupled with a quantitative analysis which measures the impact of customer trust on consumer buying intentions.

### Case Study of Tata Nano

Tata Motors Ltd. is India's largest automobile company. Its latest passenger car Nano remains a modern-day symbol of India's ingenuity. This was an outcome of a deep understanding of economic stimuli and customer needs, and the ability to translate them into customer-desired offerings through leading edge R&D.

The seeds of this innovation were sowed in the mind of Ratan Tata, chairman of Tata Motors Ltd. when he saw a family of four crammed on a two-wheeler on wet roads of a rainy day. The child was standing in front, with his mother riding pillion while holding a baby. He felt a strong need to provide them with a secured and affordable vehicle- a people's car. Tata primarily aimed to serve the needs of those at the bottom of the pyramid who so far could afford a two-wheeler but not a car. This car was proposed to be less than half the price of the cheapest car available in India and, indeed, anywhere in the world. This way Tata envisioned to create a 'new market for cars which does not exist', making them accessible to India's middle classes growing at around 9 per cent a year. However, there was a huge difference in switching cost from a two wheeler to a car. The announcement of Tata Nano aimed to reduce this gap, with a target sales price of 2,200 USD, thereby making the shift easier for the bottom of pyramid.

Tata's dream project took off in the year 2003. The idea was to develop an innovative, attractive and cost-effective means of transportation for the underprivileged while balancing the customer's expectations and meeting the regulatory requirements. Innovation with new market segment requires major process reengineering which has to be accepted by the customers. Tata tried new design of Nano to keep the cost low. In order to ensure a spacious interior, lower weight and low costs, engine was strapped in the car's rear, with front wheel drive and the petrol tank to the front. This made the car more low-cost, efficient and compact. A lot of fibre and plastic were used instead of steel to keep the weight of the car low. No radio, power windows, air conditioning, anti lock brakes, air bags, remote locks or power steering were part of the car.

Rear wheel drive had manually actuated 4-speed trans-axle that gives the car better fuel efficiency. It had strong wheel bearing to drive the car at 72kmph. Finally, Nano was produced in three variants- standard and two deluxe models with AC.

Tata Nano was built without any compromise on quality, emission and safety standards. A development, which signifies a first for the global automobile industry, Nano brought the comfort and safety of a car within the reach of thousands of families. Nano was launched with great fanfare at the 9th Auto Expo held in New Delhi fulfilling the dream of millions of Indians of owning their own car. At its price, it was quite proportionate and well styled to comfortably seat four adults.

The major targeted segment was bottom of the pyramid with two wheelers, who aspired for a four wheeler which was beyond their pocket. As this product offering was first of its kind for this segment, easy acceptability could come when communication was designed specifically for them. Hence, novel ways of marketing which were never associated with passenger car vehicles were devised. In order to keep their communication campaign innovative yet cost-effective, Tata advertised through print medium and radio. Other strategies included- online Nano games, Nano chat rooms, Nano conversations on facebook, orkut and blogs, Nano pop-ups on major websites launching Nano merchandise like baseball caps, key chains, and T-shirts etc. The Tata group also channelled marketing efforts through, Tata Sky (satellite television) where new customers could obtain a special 20 per cent discount on their satellite connection by submitting their booking proof for Tata Nano at any authorized dealer. In addition, Westside, the Tata group-owned lifestyle retail chain, advertised Nano through text messages to customers (Wells, 2010).

The distribution network of Tata Nano was also very different from contemporary one. Lower income customers were apprehensive and hesitant to walk into large Tata Motors Ltd. showrooms. It was also sold through its own retail and electronics megastore (Westside and Chroma) outlets as well as auto dealerships. In order to get substantial demand from the remotest corner of the country, the sale of form for booking were facilitated through 18 preferred banks / Non-Banking Financial Company (NBFCs). New insurance schemes were co-designed with five partner insurance companies to enhance the sales and service network for better reach and service to the customers. The prospective customers had to book Nano with INR 3,500 with the banks. From the bookings a lottery system was adopted to select customers for delivery of cars.

Taking cues from customers' responses they introduced some better strategies to woo more customers. They launched low-key access sales points called 'F Class' showrooms that display just one car. In order to reach smaller towns, they have set up special Nano access points to experience and test-drive the car (ET Bureau, 2010). Zero per cent finance schemes with 4-year extended warranty and a maintenance contract of INR 99 was introduced to boost customer confidence (Philip and Chauhan, 2010).

Even with small hiccups of Nano catching fire, the company gained back the confidence of customers by installing additional safety features as a retrofitting exercise in the 70,000 Nanos already sold. The company has made a tie up with value retailer Big Bazaar to bring touch and feel experience to the customer.

This case study helps to understand the macro picture of how innovation of Tata Nano helped create new market for bottom of the pyramids. In order to find out whether this innovation is sustainable it was important to explore its acceptability within its targeted segment. Thus, an empirical survey was carried out that helps in identifying the major attributes that make it attractive for its targeted customers. The next step is to find and their impact on customer's intent to purchase. The details of the survey are given in the subsequent section.

**Empirical Survey**

The unit of analysis in this study is a prospective customer of Tata Nano. The purpose of the survey was to find whether its potential customers would be interested in buying Tata Nano. For this it was important to understand the various attributes that would make Nano appealing to them. These attributes were found using deductive technique. In-depth interviews were conducted on respondents to understand the qualities that they associate with Nano and the factors that would prompt them to purchase Nano. Selection of participants ensured that they were 'appropriate' opinion leaders with well-developed views on the research topic (Minichiello, Aroni, Timewell, & Alexander, 1995).

The interview transcripts were content analysed and several categories were formed based on their frequency with which they appeared in the interviews. The findings were categorized into performance, brand equity, value for money, safety concerns, service, comfort, aesthetics and perception of others/status symbol (eight categories). Items were developed for each category. Some items were adapted from Jarvenpaa et al., (2000); Doney and Cannon, (1997). Since the dependent variable was intent to purchase, items were also developed for the same. After a check on content validity by two subject matter experts, items were modified/ deleted. This resulted in a final questionnaire with 33 items on attributes of Nano and seven on intent to purchase. A five-point Likert scale with end points of 'strongly disagree' and 'strongly agree' was used. A survey was conducted in four major cities of Uttar Pradesh, a heavily populated state of India. Of the 150 questionnaires distributed, 123 were returned out of which 108 were complete in all respects. The questionnaire also included the demographic profile of the respondents as these factors are likely to influence their buying behavior. The profile of respondents is given in **Table 1.**

# Table 1

## Respondents Profile

| Respondent profile | Percent |
|---|---|
| Profession | |
| Salaried | 38.7 |
| Business | 29.7 |
| Home maker | 19.9 |
| Student | 11.7 |
| Age | |
| 15-25 years | 13.6 |
| 26-30 years | 27.0 |
| 31-35 years | 36.0 |
| 36-40 years | 9.0 |
| 41-50 years | 12.6 |
| Above 50 years | 1.8 |
| Gender | |
| Male | 60.3 |
| Female | 39.7 |
| Educational Quali cation | |
| High school | 1.8 |
| Intermediate | 7.2 |
| Graduate | 30.6 |
| Postgraduate | 58.6 |
| Any other | 1.8 |
| Overall Work Experience | |
| NIL | 9.9 |
| < 1 year | 9.0 |
| Between 1-5 years | 10.8 |
| Between 5-10 years | 19.8 |
| Between 10-15 years | 36.9 |
| Above 15 years | 13.6 |
| No. of members in your family | |
| 1 | 3.6 |
| 2 | 10.8 |
| 3 | 20.7 |
| 4 | 35.2 |
| 5 and above | 29.7 |
| Income per annum | |
| Less than INR 1 lakh p.a. | 14.4 |
| Between INR 1 lakh-3 lakhs p.a. | 30.6 |
| Between INR 3 lakhs-5 lakhs p.a. | 23.4 |
| Between INR 5 lakhs-10 lakhs p.a. | 18.1 |
| Greater than INR 10 lakhs p.a. | 13.5 |
| Two Wheeler owners | 85.8 |
| Car owners | 12.3 |
| House owners | 56.7 |
| Air-conditioner owners | 11.3 |
| Personal computer owners | 15.2 |
| Color TV owners | 83.7 |

According to Table 1, nearly 70% of the respondents are salaried and businessmen. Nearly 60% of the respondents fall into the age group of 26 to 35 years. Table 1 also points out that respondents are from different income backgrounds with significant majority of the respondents in income group of INR 1-5 lakh p.a. Education background shows maximum percentage of respondents to be post-graduates. It also depicts majority of the respondents having more than 4 members in family. The respondents' demographic profile clearly brings out characteristics of Indians falling in the middle class range. This is the burgeoning class that has evolved with the changing economic scenario and is willing to buy passenger car.

An exploratory factor analysis (EFA) using principal component analysis was carried out to examine the factor structure. The commonly recommended method of orthogonal varimax rotation with Kaiser Normalization and MINIEIGEN criteria was used. KMO and Bartlett's Test of sphericity was found to be significant i.e. above 0.2 for both the constructs. This signifies that some correlation existed in order to proceed for factor analysis. Next, the communalities were evaluated and only those items with communalities above 0.5 were retained. Items that contributed least to the overall internal consistency were the first to be considered for exclusion. The item inter-correlation matrix showed that there were no items that were negatively correlated to other items within scale. All the items had their correlation value between 0.3 and 0.8 with at least one item of the scale.

Exploratory factor analysis resulted in the deletion of 20 items from attributes of Nano and 1 from intent to purchase due to cross loadings and loadings below 0.4 **(see Table 2 & 3)**. The attributes of Nano resulted in clear four factors out of which three (corporate brand equity, aesthetics, value for money) were same as derived from personal interviews while the fourth one had an item each from the categories performance and service. We relabelled it as reliability. Out of the five items that emerged for Tata, the corporate brand, one item 'Tata Nano has appealing range of colors' was specific to the product brand. Corporate brand equity emerged as the strongest factor with the variance of 31 per cent followed by aesthetics, value for money and reliability. Similarly, the construct of intent to purchase resulted in two factors, which were labelled as decision making behaviour and information seeking behaviour. The variance exhibited by these constructs was found to be 61.4 per cent and 66.6 per cent respectively. The coefficient alphas for the attributes and intent to purchase dimensions are 0.80, and 0.78 respectively, which are above the threshold of 0.70 recommended by Nunnally (1978).

## Table 2

## Factor structure of 'Attributes of Nano'

### Rotated Component Matrixa

| | Component | | | |
|---|---|---|---|---|
| | Corporate brand equity | Aesthetics | Value for money | Reliability |
| Tata is a reliable brand. | .804 | | | |
| I trust Tata company to keep its customers' interests in mind. | .800 | | | |
| The quality of different cars produced by Tata is consistently high. | .742 | | | |
| Tata Nano has appealing range of colors. | .631 | | | |
| I feel assured of car delivery within the specied time. | .485 | | | |
| I nd the car to be attractive. | | .803 | | |
| I nd the car to be comfortable for me and my family. | | .745 | | |
| Tata Nano is beautifully designed. | | .674 | | |
| Tata Nano suits my budget. | | | .750 | |
| It is easy to drive in heavy trafc. | | | .728 | |
| This car can be counted on to show good performance. | | | .710 | |
| I am aware that it has service centers even in smaller cities. | | | | .838 |
| I am condent to drive this car. | | | | .696 |

Extraction Method: Principal Component Analysis.
Rotation Method: Varimax with Kaiser Normalization.
aRotation converged in 6 iterations.

## Table 3

## Factor structure of 'Intent to purchase'

## Rotated Component Matrixa

|  | Component | |
|---|---|---|
|  | Decision making behavior | Information seeking behavior |
| I intend to buy Nano in the near future. | .889 |  |
| I would consider devoting my time or money to this car. | .885 |  |
| Perception of others about Tata Nano would inuence my decision to buy the car. | .477 |  |
| I intend to collect more information about this car. |  | .873 |
| I have already taken opinion of my friends and relatives to buy this car. |  | .806 |
| Latest advertisement has motivated me to buy Tata Nano. |  | .654 |

Extraction Method: Principal Component Analysis.
Rotation Method: Varimax with Kaiser Normalization.
a Rotation converged in 3 iterations.

## Table 4

## Regression Summary

| Model | R | $R^2$ | Adjusted $R^2$ | Standard error of the estimate | Durbin Watson |
|---|---|---|---|---|---|
| 1 | 0.650 a | 0.423 | 0.417 | 6.4124 | 1.786 |
| **ANOVA** | | | | | |
| Model | Sum of Squares | df | Mean Square | F | Significance |
| 1   Regression | 32.198 | 1 | 32.198 | 78.306 | 0.000 |
|      Residual | 43.997 | 107 | 0.411 | | |
|      Total | 76.195 | 108 | | | |
| [a] Predictors: (Constant), ATB | | | | | |
| [b] Predictors: (Constant), ITP | | | | | |
| **Coefficients [a]** | | | | | |
|  | Unstandardized Coefficients | | | | Significance |
| Model | β | Std. Error | β | t | |
| (Constant) | -0.258 | 0.370 | 0.650 | -0.698 | 0.487 |
| ATB | 0.932 | 0.105 | | 8.49 | 0.000 |
| Dependent Variable: ITP | | | | | |
| [a] Predictors: (Constant), ATB | | | | | |

When we examine **Table 4**, which reflects the attributes of Nano (ATB) as the predictor of intent to purchase (ITP), we find that the adjusted R2 shows a significant value of 0.42, implying that attributes of Nano make a moderate difference to customers' intent to purchase Nano. Thus 42 per cent of the variance in intent to purchase is explained by variance in attributes. ? value shows that for every unit rise in attributes, intent to purchase is increased by 65per cent. Further, when we refer to Table 4, we find that the ANOVA between attributes of Nano and the intent to purchase is significant: (df = 107, 108, f = 78.306, p = 0.000). Further to test the statistical significance of ? value we examine the t values. The t value for attributes (8.849) is significant for intent to purchase.

## CONCLUSION

### Findings

The findings reveal that corporate brand equity, aesthetics, value for money and reliability are the major attributes of Tata Nano that makes it attractive for its targeted customer. Bottom of the pyramid is very price-sensitive. They cannot afford to misplace their money in non-reliable product offerings. They care about branded products, as they associate them with quality and reliability. Existence of Nano is not only because of the efforts put in reengineering and new market strategies, but was majorly due to high corporate brand equity of brand 'Tata' that made their efforts look reliable and show value for money for this segment.

A surprise finding was emergence of aesthetics as second most influential factor for Nano in terms of its design, attractiveness and comfort. The reason could be that because of its low price tag, everyone imagined it to be an ugly looking boxy hatch. Surprisingly, it was quite proportionate and well styled. Its mono box design generated a lot of interior space which could comfortably seat four adults. It proved its mettle by asserting itself as a value for money product and a reliable product. Since the service centres were in small towns also, it gave them the confidence having help at hand. During regression analysis a moderate relationship was found between the attributes of the car and intent to purchase. This implies that even though these attributes given an identity to Nano, they do not majorly impact customers' intent to purchase. The reason could be that the customers do not want to associate themselves with a people's car that is specifically positioned for the bottom of the pyramid.

## IMPLICATIONS

A key implication of this paper is that the organizations must adapt 'blue ocean strategy' (Kim & Mauborgne, 2005) whereby organizations can generate high growths and profits by creating new demand in an uncontested market space. The empirical findings suggest several important academic and practical contributions along with several applications for the research. Its academic contribution is to offer a significant advancement to the current literature of new market creation and innovation. For the practitioners it implies that they should highlight these attributes in their advertising and communication strategies. Instead of emphasizing on low price tag, value for money should be highlighted so that stigma of buying a people's car is removed.

Limitations and direction for future research

One of the limitations of the study is the geographical extent of the survey. The present findings are therefore indicative rather than conclusive. There also exists some possibility of response biases occurring due to differences in perception, attitude, and behavior. The paper concludes with directions for future research. The innovative product chosen is specifically a passenger car-Nano. Further research possibility lies in other product categories. By increasing the geographical extent the real picture of acceptability of Nano could emerge. The impact of various advertising strategies used by Nano on the intent to purchase could be analysed to estimate their efficacy.